\documentclass{kluwer}

\def\sun{\hbox{$\odot$}}

\newcommand{\solarmass}{M$_{\sun}$\,}
\newcommand{\solarmassyr}{\solarmass .yr$^{-1}$}

\begin{document}
\begin{article}
\begin{opening}
      
\title{Physical and chemical structure of the solar type protostar NGC1333-IRAS4}
      
\author{S. \surname{Maret}\email{sebastien.maret@cesr.fr}}
\institute{Centre d'Etude Spatiale des Rayonnements, CESR/CNRS-UPS, BP
  4346, F-31028 - Toulouse cedex 04}
      
\author{C. \surname{Ceccarelli}}
\institute{Observatoire de Bordeaux, BP 89, F-33270 Floirac}
      
\author{E. \surname{Caux}}
\institute{Centre d'Etude Spatiale des Rayonnements, CESR/CNRS-UPS, BP 4346,
  F-31028 - Toulouse cedex 04}            

\end{opening}

NGC1333-IRAS4 is a binary protostellar system (IRAS4 A and B) located
in the south part of the Perseus cloud. They have been classified as
Class 0 protostars \cite{Andre00} and are associated with molecular
outflows probed by CO and CS millimeter lines.\\ We observed IRAS4
with the \emph{Long Wavelength Spectrometer} \cite{Clegg96} on board
the \emph{Infrared Space Observatory} \cite{Kessler96} in grating
mode. These observations consist of a spectral survey of the central
position containing both IRAS4A and B, and two positions along the CO
outflow (see also Ceccarelli et al., 1999).  We detected fourteen
H$_{2}$O lines, nine CO lines, and the [OI] and [CII] fine structure
lines. No significant water and CO emission was found along the
outflow, whereas the [OI] and [CII] lines were detected with the same
intensity than on the central position.\\ The observed molecular
emission can a priori have three different origins: the outflows, a
photodissociation region (PDR), and the collapsing envelope around the
protostar. The [OI] and [CII] emission can be relatively well
explained by a PDR with a density from 10$^{4}$ to 10$^{6}$ cm$^{-3}$,
and an incident FUV field, whose flux is of the order of the average
interstellar FUV field flux. On the contrary, both CO and H$_{2}$O
emission could arise either in shocked gas (e.g. due to the impact of
the outflowing gas with the envelope) and/or in the envelope itself.\\
In the past we \cite{Ceccarelli96} computed the thermal emission of a
protostellar envelope. This model computes in a self consistent way
the thermal balance, chemistry and radiative transfer in the envelope,
in the ``inside-out'' framework \cite{Shu77}. We successfully applied
this model to IRAS4, and we were able to explain the H$_{2}$O emission
as due to an envelope collapsing towards an object whose mass is 0.6
\solarmass and accreting at 5.10$^{-5}$ \solarmassyr (see Maret et
al., in preparation, for more details).  Assuming a constant mass
accretion rate this gives an age of 12000 yr. The water abundance is
about 10$^{-6}$ with respect to H$_{2}$ in the outer cold envelope,
and it is enhanced by one order of magnitude in the central region,
where the grain mantles evaporate. The existence of such
\emph{hot-core} like region has been also claimed in the case of the
solar type protostar IRAS16293-2422, where a similar study was carried
out \cite{Ceccarelli00a,Ceccarelli00b}. IRAS16293-2422 seems to be
more massive (0.8 \solarmass) than IRAS4 and accreting at a lower
accretion rate ($3.10^{-5}$ \solarmassyr), suggesting that
IRAS16293-2422 is more evolved than IRAS4, with an estimated age of
27000 yr. This conclusion is in agreement with the relatively large
millimeter continuum observed in IRAS4, which implies a large amount
of dust surrounding this source. Finally, both the hot core like
region and the region in which CO-rich ices are predicted to evaporate
(i.e. when the dust temperature is $\sim25$ K) are larger in
IRAS16293-2422 than in IRAS4, which can explain the larger CO
depletion observed in the latter \cite{Blake95}.\\
This study also emphasizes the necessity of ground based observations,
where higher spatial and spectral resolutions are
achievable. H$_{2}$CO and CH$_{3}$OH are of particular interest as
they are among the most abundant components of the grain mantles, and
are therefore expected to evaporates in the innermost parts of the
envelope. Appropriates transitions can hence be used to constrain the
physical and chemical conditions in such innermost parts of
protostellar envelopes.

\end{article}
\end{document}